\documentclass[aps,prc,twocolumn,superscriptaddress,showpacs]{revtex4}

\usepackage{graphicx}
\usepackage{times,mathptmx}
\usepackage{amsmath}
\usepackage{amssymb}

\begin{document}
\preprint{}

 \title{Origin of the narrow, single peak in the fission-fragment mass distribution for $^{258}$Fm} 

\author{Takatoshi Ichikawa}%
\affiliation{RIKEN Nishina Center, RIKEN, Wako, Saitama 351-0198, Japan}
\author{Akira Iwamoto}
\affiliation{Japan Atomic Energy Agency, Tokai-mura, Naka-gun, Ibaraki
319-1195, Japan}
\author{Peter M\"oller}
\affiliation{Theoretical Division, Los Alamos National Laboratory, Los Alamos, New Mexico 87545, USA}
\date{\today}

\begin{abstract}
 We discuss the origin of the narrowness of the single peak at
 mass-symmetric division in the 
 fragment mass-yield curve for spontaneous fission of $^{258}$Fm.
 For this purpose, we employ the macroscopic-microscopic model, and
 calculate a potential-energy curve at the mass-symmetric 
 compact scission configuration, as a function of the fragment mass
 number, which is obtained from the single-particle wave-function densities.
 In the calculations, we minimize total energies by varying the deformations of
the two fragments, with constraints on the mass quadrupole moment and keeping the
 neck  radius zero, as a function of mass asymmetry.
 Using the obtained potential, we solve the one-dimensional
 Schr\"odinger equation with a microscopic coordinate-dependent inertial mass 
 to calculate the fragment mass-yield curve. 
 The calculated mass yield, expressed in terms of the microscopic mass
density, is consistent with the extremely narrow
 experimental mass distribution.
\end{abstract}

\pacs{24.75.+i, 27.90.+b}

\maketitle
In spontaneous fission the fragment mass-yield distributions
change abruptly from a double-peaked, broad, mass-asymmetric
distribution for $^{256}$Fm to a single-peaked, very narrow, symmetric
distribution for
$^{258}$Fm ~\cite{DCHoff80,Hulet86,Hulet89}. In addition, in $^{258}$Fm
the kinetic-energy distribution can be expressed as a sum of a low-energy
and a high-energy component, whose mean energies differ by
about 35 MeV\@.
The mechanism behind this phenomenon, called bimodal fission, is
the strong nuclear shell effects that appear when symmetric division into two
fragments which both are near the doubly magic nucleus
 $^{132}$Sn becomes possible.
The experimental observations of the
sudden emergence of  a mass-symmetric division
near $^{258}$Fm has always been assumed
to be due to an 
 emergence in fission potential-energy surfaces 
of  a fission path, strongly 
stabilized by fragment shell
effects~\cite{DCHoff80,Zhao99,Zhao00,Britt84,DCHoff95}. However, no
such intuitive picture of the mechanism behind the narrowness of
the symmetric mass distribution peak has been advanced,
nor has a convincing quantitative calculation  explaining the extremely narrow FWHM 
 been presented.     

So far, theoretical investigations have mainly focused on 
obtaining the transition point between competing fission modes
near $^{258}$Fm by calculating the potential-energy surface
versus various chosen sets of deformation coordinates.
In fact, theoretical models, such as the macroscopic-microscopic
model \cite{Moller87,Pashk88,Moller89,Cwiok89,Moller94,Moller01}, the constrained 
Hartree-Fock+BCS (HFBCS) model~\cite{Bonneau06,Staszczak07}, and the
constrained Hartree-Fock-Bogoliubov (HFB) model~\cite{Warda02,Dubray08},
have to a varying degree of success described such a shell-stabilized path,
and the transition point between asymmetric and symmetric fission modes
near $^{258}$Fm, that is,  the emergence here of 
path leading to mass-symmetric divisions with compact scission
configurations, referred to as the compact symmetric path.  
Through these investigations the 
energy-minimum path leading to high-kinetic-energy, 
symmetric fission has been well established, but the structure of the
{\it potential valley} along this  path 
and plausible mechanisms 
behind the extremely narrow FWHM 
have been less extensively studied.

In this paper, we calculate quantitatively the mass distribution of 
$^{258}$Fm  in the compact symmetric valley.
The calculations are based on studies of the dynamics 
of the  zero-neck-radius  scission configuration in the mass-asymmetry
shape degree of freedom~\cite{Vand73}.
We also impose spherical fragment shapes, 
which leaves us with
mass asymmetry as the only collective
coordinate.
This approach implies that  the
mass distribution originates from zero-point vibrations or thermal fluctuations
in the mass-asymmetry degree of freedom at scission.
In the specific case of $^{258}$Fm considered here,
it is a reasonable assumption and approximation,
since in spontaneous fission the exit point after barrier penetration
is approximately the zero-neck scission configuration.

We  use the macroscopic-microscopic model~\cite{Bols72,Moller95} constrained
to shapes with zero neck radius.
Using a macroscopic-microscopic model, slightly different
from our implementation, Pashkevich failed to find any
significant difference between the 
 curvatures of the compact and elongated symmetric
valleys for $^{264}$Fm~\cite{Pashk88}. Furthermore, in his cranking-model
analysis of the mass distribution widths associated
with the zero-point oscillations in his asymmetry degree of freedom he
obtained similar mass yield distribution widths in both valleys, approximately
consistent with the narrow distribution observed experimentally
for $^{258}$Fm. His dynamical study has significant similarities with
our study here.
However, Pashkevich characterized fragment mass numbers by the 
asymmetry of the homogeneous macroscopic volume defined
by the parametrization of the nuclear surface.
Here we use microscopic densities to
characterize the mass distribution, which gives
very different mass-distribution curves close to magic numbers.
As we discuss in detail below, we propose it is the shell gaps
that restrict fluctuations in the microscopic mass-asymmetry degree of freedom.

We use the three-quadratic-surface 
parametrization~\cite{Nix68,Nix69} to describe macroscopic nuclear
shapes in a five-dimensional deformation space. The shape degrees
of freedom are a quadrupole-moment parameter
$q_2$, a neck parameter $\eta$,  left- and right-fragment
deformation parameters, $\epsilon_1$ and $\epsilon_2$, respectively, and
a mass-asymmetry parameter $\alpha_{\rm g}$.
The parameter $q_2$ is the dimensionless quadrupole moment in  units of
$3ZR_0^2/4\pi$ (e$^2$b), where $Z$ is the proton number and $R_0$ is the
nuclear radius.
The parameter $\eta$ varies  from 0 to 1.
Scission, with zero neck radius,  
corresponds to $\eta=0$.
The parameter $\epsilon$ is
the Nilsson perturbed-spheroid parameter. 
Near scission we have to a very good approximation
$\alpha_{\rm g}=(M_1 - M_2)/(M_1 + M_2)$,
where $M_1$ and $M_2$ are the volumes of the left and right nascent
fragments, respectively. 
The microscopic single-particle potential is calculated by folding a Yukawa
function over the macroscopic shape or ``sharp-surface generating volume''
\cite{Bols72}.

\begin{figure}[t]
\includegraphics[keepaspectratio,width=\linewidth]{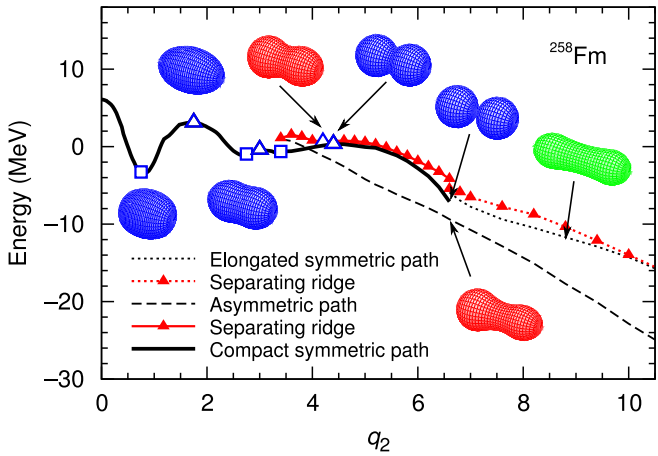}%
\caption{\label{fig1} (Color online) Potential energy as a function of
 the dimensionless quadrupole moment parameter $q_2$.
 The open squares and triangles are
 the minima and optimal saddles obtained by the immersion method
 in the five-dimensional potential-energy surface, respectively.
 The solid and dotted lines are the valleys leading to 
 compact and elongated symmetric fission.
 The dashed line is the valley leading to the asymmetric fission.
 The solid and dotted lines with triangles represent the separation ridge between
 the compact symmetric and the asymmetric valleys and the separating
 ridge between the elongated symmetric and the asymmetric valleys, respectively.
  }
\end{figure}

\begin{figure}[t]
\includegraphics[keepaspectratio,width=\linewidth]{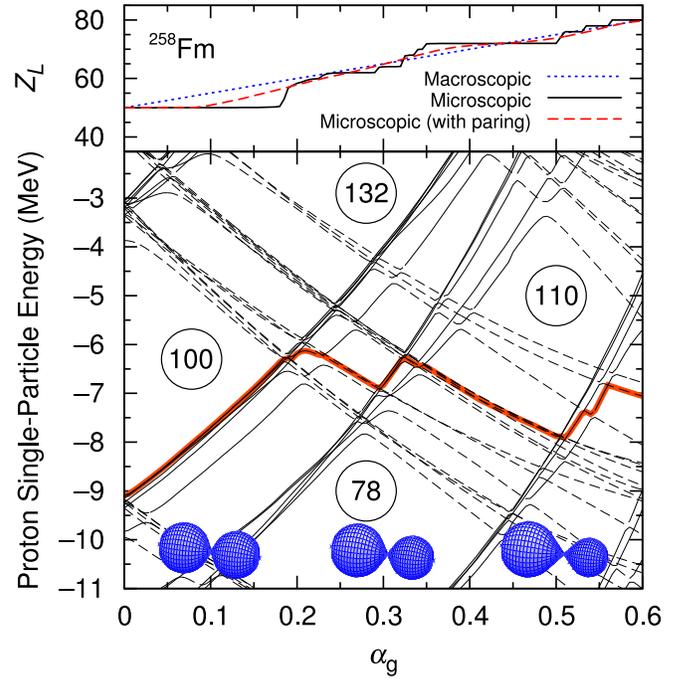}%
\caption{\label{fig2} (Color online) (Upper panel) Proton number of 
 left fragment as a function of the mass-asymmetric parameter
 $\alpha_{\rm g}$. The 
 solid line is the  total proton density obtained as a sum over occupied  
wave functions in the left fragment.
 The
 dashed line is the total proton density with pairing taken into account.
The dotted line is the macroscopic left-fragment density.
 (Lower panel) Nilsson diagram for proton
 single-particle levels at scission configurations in the compact
 symmetric valley as a function of the mass-asymmetric parameter
 $\alpha_{\rm g}$.
 The solid and dashed lines are the energy levels whose wave function
 localizes in the right- and left  fragments, respectively. 
 The bold-gray (red) line is the Fermi level.
 }
\end{figure}

\begin{figure}[t]
\includegraphics[keepaspectratio,width=\linewidth]{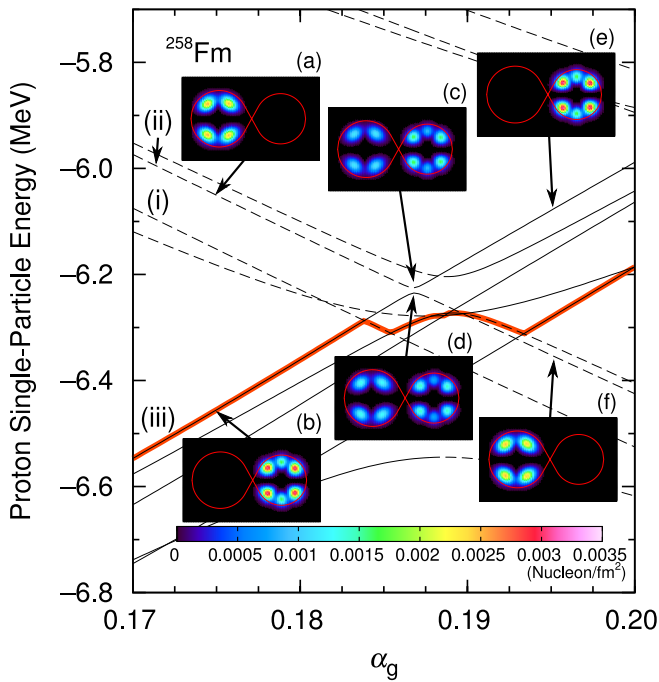}%
\caption{\label{fig3} (Color online) Enlargement of part of Fig.~2 with
 density plots of a few specific wave functions discussed
in the text. In the insets, the
 corresponding macroscopic nuclear shape is given by the solid line.
}
\end{figure}

To study the properties of the compact, mass-symmetric valley, we calculate the
five-dimensional potential-energy surface for $^{258}$Fm and analyze it by use of
the immersion method.  
Details of the calculation are given in Ref.~\cite{Moller08}.
The parameters correspond to FRLDM(2002)~\cite{Moller04}.
For simplicity, we calculate the pairing effect based on the BCS
model~\cite{Moller92}. Since we mainly consider scission or 
near-scission shapes, we can ignore the shape dependence of the Wigner
term. 
We calculate the potential energies at
$41\times20\times15\times15\times35$ grid points  for $Q_2$, $\eta$,  
$\epsilon_1$, $\epsilon_2$, and $\alpha_{\rm g}$, respectively.

Figure 1 shows some main structures identified by immersion techniques in
the calculated  potential-energy surface for
$^{258}$Fm as a function of $q_2$. 
 The open squares and triangles are  minima and optimal saddles
between minima, 
respectively. The energy of the saddle point on the path to compact
 symmetric fission is 0.34 MeV, while that of the asymmetric fission is
 0.58 MeV. 
 The scission point for the compact symmetric fission is at $q_2=6.5$,
 $\epsilon_{1,2}=0$, and $a_{g}=0$. 
 In the figure, we show the compact symmetric and the asymmetric 
 valleys, denoted by a solid and a dashed line, respectively. Those are
 separated by a ridge, denoted by a solid line with triangles. 
 We also find a path leading to elongated mass-symmetric divisions,
 denoted by a dotted line, but its separating ridge to the 
 asymmetric valley, denoted by the dotted line with triangles,
 vanishes at around $q_2=10.0$, indicating that the scission point of 
 this path strongly depends on the dynamics after going through this valley.
 
 Before we discuss the dependence of the potential energy on mass asymmetry at
the compact scission configuration we need to discuss the 
 relation between fragment mass numbers and single-particle
 energy levels. For this purpose, we calculate the proton density for each
 single-particle state and the total proton number of the left fragment
 as a function of $\alpha_{\rm g}$.
 For $\alpha_{\rm g}>0$, the volume of the left fragment is
 greater than the right fragment.
 We take $\epsilon_{1}$ and $\epsilon_{2}$ to be 0 in
 the calculation.  
 If we maintain our restriction to  axially symmetric shapes, the proton density
 for the $i$-th single-particle state is given by
 $|\psi_i(\rho,z)|^2$, where $\psi$ is the single-particle wave
 function in the cylindrical coordinate system. The single-proton occupation
 probability in left
  of two nascent fragments is thus obtained by 
 \begin{eqnarray}
   n_i=2\pi\int_{0}^{\rho_{\rm max}}\int_{z_{\rm min}}^{z_{\rm neck}}\rho |\psi_i(\rho,z)|^2dzd\rho,
 \end{eqnarray}
 where $z_{\rm neck}$ is at a macroscopic neck radius  of 0 fm.
 The values $\rho_{\rm max}$ and $z_{\rm min}$ (negative) are set sufficiently
large that $|\psi_i(\rho,z)|^2$
 becomes negligible outside the integration intervals. 
 The total proton number $Z_{\rm L}$ is the sum of the single-proton
 occupation probability from the lowest level to the Fermi level.
 We also calculate the total proton number taking into account the
 pairing effect, $Z_{\rm L}^{\rm (BCS)}$, given by $Z_{\rm L}^{\rm (BCS)}=\sum
 v_i^2 n_i$, where $v_i^2$ is the occupation probability calculated
 using the BCS pairing model.
 
 We use two density concepts when we discuss the mass asymmetry
in our study. One definition is based on calculating the
mass asymmetry from the single-particle densities, the other
from the asymmetry of the homogeneous volume defined
by the parametrization of the nuclear surface, denoted 
``microscopic'' and ``macroscopic'', respectively.

 Figure 2 shows important  features of our results.
 The solid and dashed lines in the upper panel of Fig.~2 show the total
number of protons
 without and with the pairing effects, $Z_{\rm L}$ and $Z_{\rm
 L}^{(\rm BCS)}$, respectively. 
 The lower panel of Fig.~2 is a Nilsson diagram versus $\alpha_{\rm g}$.
  The thick gray (red) line is the Fermi level.
The energy levels are plotted dashed when more than half the density
is in the left part of the potential, solid otherwise 
 [see the insets from (a) to (f) in
 Fig.~3].
 The upper panel shows that in the absence of pairing
 the total proton number remains constant at
 $Z_{\rm L}=50$ in the interval  $ 0 \leq \alpha_{\rm g} \leq 0.18. $ 
 Just below $\alpha_{\rm g}=0.20$,  $Z_{\rm L}$  suddenly jumps to 
 about 60. One can also see such discontinuities at large $\alpha_{\rm
 g}$. Those discontinuities occur where  single-particle shell
 gaps change, that is they  coincide with the level
 crossing points in the Nilsson diagram.  It is here the
 downward-sloping states whose wave functions are localized in the left fragment 
 come below the Fermi level and can be populated with particles transferring
from the right fragment.
 These sudden features are smoothed out when we include the pairing correlation.

 In order to see more clearly how $Z_{\rm L}$ increases 
we show in Fig.~3 an enlargement of the Fermi-surface
region of Fig.~2, with
some wave-function densities inserted. 
In the figure, the downward-sloping state denoted by the line (i),
coming from above the Fermi level, dives below 
 the Fermi level. In the case we have here with paired particles,
protons can  transfer from the right fragment into
this previously unoccupied state in the left fragment, which leads to an 
 increase of $Z_{\rm L}$. In contrast, in the case the particle number is
odd then the odd particle can not transfer to a level
of different $\Omega$, in a sudden shape change. Rather,
the odd particle remains in  its level until another level
with the same $\Omega$ quantum number is encountered,
giving rise to a ``specialization energy'' and an increase
in barrier height. This is the mechanism
behind the long spontaneous fission half-lives of odd nuclei \cite{Newton55:a,Rand73:a}.
When a downward sloping level encounters an upward-sloping level
of the same quantum number $\Omega$ when it crosses the Fermi
surface, then  a repulsion between the single-particle states occurs.
 This is the case for the states denoted by  (ii) and (iii), which
 have the same quantum number, 
 $\Omega=5/2$, 
 but their wave functions localize on the right- and left fragments,
 respectively [see the insets (a) and (b) in Fig.~3]. 
The transfer mechanism in the even, paired system is similar
to case (i), but in the single-particle picture
it appears different because when
these states come close to each other the states mix (``level repulsion''),
in this case near $\alpha_{\rm g}=0.185$
 [see the insets (c) and
 (d)].  After the mixing region,  the localizations of 
 the wave functions are interchanged [see the insets (e) and
 (f)], and the lower level remains below the Fermi level, but
with a new set of quantum numbers.
 This  microscopic mechanism is responsible for the increase in $Z_{\rm L}$
with increasing $\alpha_{\rm g}$.

\begin{figure}[t]
\includegraphics[keepaspectratio,width=\linewidth]{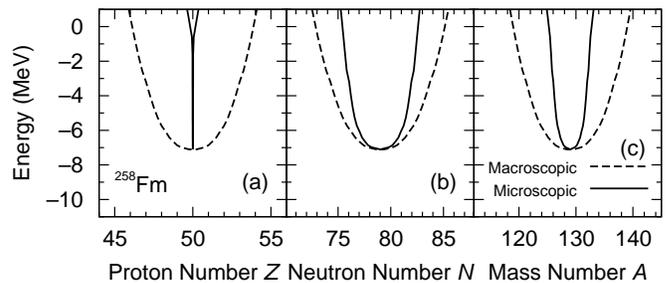}%
\caption{\label{fig4} Potential-energy curves at the scission
 configuration as a function of the (a) proton, (b) neutron, and (c) total mass 
 densities of two nascent fragments, respectively. The solid and dashed
 lines denote the potential-energy curve calculated
 using the microscopic and macroscopic densities, respectively.
}
\end{figure}

\begin{figure}[t]
\includegraphics[keepaspectratio,width=\linewidth]{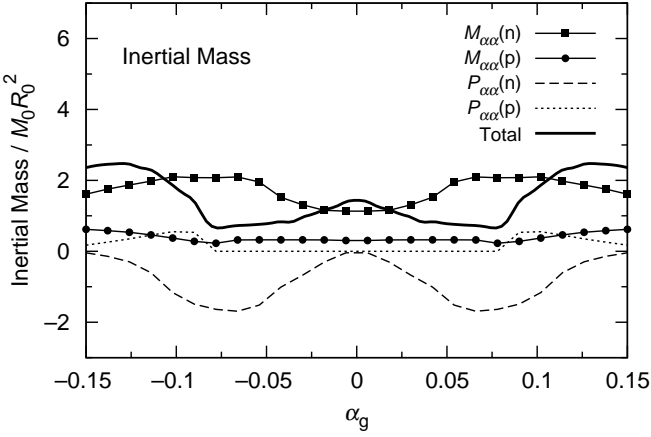}%
\caption{\label{fig5} Inertial mass in the
 mass-asymmetric direction at the compact-symmetric scission
 configuration calculated using the Inglis-Belyaev formula. The solid
 lines with squares and circles are the first term of the RHS in Eq.~3 for
 neutrons and  protons, respectively. The dashed and dotted lines
 are the second term of the RHS in Eq.~3 for  neutrons and  protons,
 respectively. The solid line is the total inertial mass  
for neutrons and protons calculated as  the sum
 of the first and second terms in Eq.~3.} 
\end{figure}
 
 As shown by the dashed line in the upper panel of Fig.~2, the
 pairing interaction smooths the change in the fragment proton number versus
asymmetry,  because it incorporates wave-function admixtures across 
the single-particle Fermi surface into the sum of the single-proton
 densities. As a consequence, if the density of states just above a shell gap
occurring at the Fermi surface is high, then  $Z_{\rm L}^{\rm (BCS)}$
 more closely tracks the macroscopic proton number.
 At large $a_{\rm g}$ this condition is particularly well fulfilled and the dashed line is
 close to the macroscopic density, denoted by the dotted line in the figure.
  However, we found that from $\alpha_{\rm g}=0$ to 0.10, $Z_{\rm L}^{\rm (BCS)}$ still
 remains constant at 50 due to the very substantial shell gap, indicating
 that proton number $Z=50$ is extremely stable.
 We will show below that this stability may to a large part be the mechanism
that leads  the very narrow fragment mass distribution.

 We now calculate the potential-energy curve at the scission configuration
 in the compact symmetric valley as a function of the fragment mass
 number. 
 In this first study we take $\epsilon_1$ and $\epsilon_2$ to be zero in
 calculating the potential energy and the inertial mass.
 We have checked this approximation by minimizing the total energy with
 respect to  $\epsilon_1$ and $\epsilon_2$ with 
 constraints on $q_2=6.5$ and a neck radius of 0 fm, as $\alpha_{\rm g}$ increases. 
 From $\alpha_{\rm g}=0$ to 0.1, we found that the minimum energy
occurs at  $\epsilon_1$ and $\epsilon_2$
  equal zero due to the large spherical shell gap, but abruptly jumps to
 other valleys near $\alpha_{\rm g}=0.1$. 
Inaccuracies due to this approximation can
therefore be expected to be fairly insignificant.
 We calculated the fragment mass number using the
 single-particle wave 
 functions with the BCS paring effect, before this jump takes place.

 Figures 4 (a), (b), and (c) show calculated  potential energies  as 
 functions of the fragment microscopic  proton, neutron, and total mass densities,
 respectively.
 These results are given by the solid lines.
 As a comparison, results versus the macroscopic density are also
 given, displayed as dashed lines.
 The potential-energy curves versus the microscopic density rise  much
 more steeply than the curves  plotted versus the macroscopic density.  
 In particular, the result versus the proton density depends
drastically on the choice of density variable.
We therefore expect that the physical origin of the narrow mass distribution is
the large shell gap. Below we investigate this hypothesis  through a
 quantitative study.

\begin{figure}[t]
\includegraphics[keepaspectratio,width=6cm]{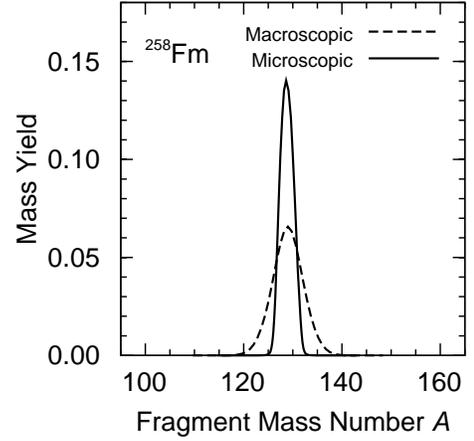}%
\caption{\label{fig6} Calculated mass-yield curves for
compact fission of $^{258}$Fm.
 The solid and dashed lines are based on the 
 microscopic and macroscopic densities, respectively.}
 \end{figure}

 In order to calculate the mass distribution in 
 the compact symmetric mode, we calculate the zero-point vibration
 corresponding to  the calculated  potential-energy curves displayed in
 Fig.~4, by solving the  one-dimensional Schr\"odinger equation in terms
 of $\alpha_{\rm g}$ with a coordinate-dependent inertial mass~\cite{LIC73}.
 For the compact symmetric valley, we expect the zero-point vibration to be dominating,
because the observed excitation energy of the fission fragments of the nearby
$^{260}$Md  in compact, symmetric fission was extremely low~\cite{Wild90}. 
 The Schr\"odinger equation thus reads 
  \begin{eqnarray}
   \left[-\frac{\hbar^2}{2\sqrt{B}}\frac{\partial}{\partial
    \alpha_{\rm g}}\frac{1}{\sqrt{B}}\frac{\partial}{\partial
    \alpha_{\rm g}}+V(\alpha_{\rm g})\right]\psi(\alpha_{\rm g})=E\psi(\alpha_{\rm g}),
  \end{eqnarray}
  where $\psi$ is the wave function and $B$ is the inertial mass in the
 mass-asymmetric direction $B_{\alpha \alpha}$.

 For the inertial mass, we employ the Inglis-Belyaev
 formula~\cite{BRA72,Rand76}, given by 
  \begin{eqnarray}
   B_{\alpha \alpha}=2\hbar^2\left[\sum_{\nu \mu}\frac{|<\nu|\partial
			      H/\partial
			      \alpha|\mu>|^2(u_{\nu}v_{\mu}+u_{\mu}v_{\nu})^2}{
			     (E_\nu+E_\mu)^3 }\right]+P_{\alpha \alpha},
  \end{eqnarray}
 where $H$ is the single-particle Hamiltonian, $v_\mu$ and $u_\mu$
 are the BCS occupancy and vacancy amplitudes, and $E_{\mu}$ is the energy of the
 quasi-particle state $|\mu>$.
 The term $P_{\alpha \alpha}$ gives the contribution from couplings to
 the pairing vibrations.
 We use the finite-difference method to calculate
 $<\mu|\partial H(\alpha)/\partial \alpha|\nu>$.
Figure 5 shows the calculated inertial mass  in  units of $M_0 R_0^2$,
where $M_0=931.50$ MeV is the atomic mass unit. The solid lines with
squares and circles show  the evaluated first term in 
the right hand side  of
Eq.~3 for the neutrons and the protons, respectively. The dashed and
dotted lines show the evaluated  $P_{\alpha \alpha}$ for  neutrons and  protons,
respectively. The solid line is the sum of those four terms.

 We calculate $\psi(\alpha_{\rm g})$ in Eq.~2 using the finite-difference method.
The calculated zero-point energy  is $-6.05$ MeV.
 The calculated  $\psi(\alpha_{\rm g})$, which is the macroscopic density
 amplitude, is converted into a mass yield function $Y(A)$ in 
 terms of fragment microscopic densities through,
 $Y(A)=C|\psi(A)|^2$, where $C$ is a renormalization factor. 
 We chose $C$ so that the total area of $Y$ is equal to 0.5,
 because the component originating from the 
 compact symmetric valley is about 50\% of the total yield. This 
has been  estimated from the two-mode 
 analysis of the experimental total kinetic-energy distribution~\cite{Hulet89}.
 The structure of our calculated potential-energy
surface  is consistent with this assumption, because the calculated
 height of the saddle point leading to the compact symmetric valley is  
 comparable to that of the asymmetric valley. See also Refs. \cite{Moller94,Moller08}.

\begin{figure}[t]
 \includegraphics[keepaspectratio,width=\linewidth]{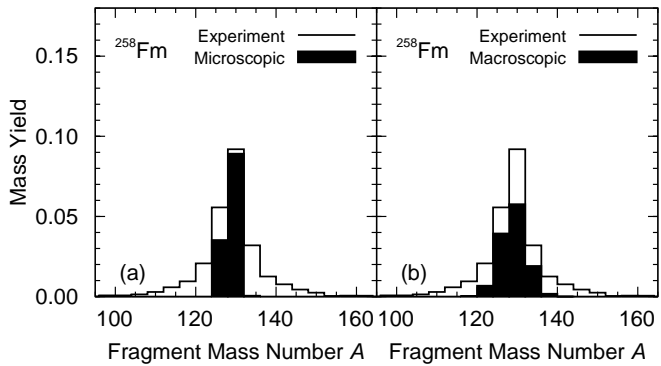}%
 \caption{\label{fig7}  Mass yields plotted in histogram form.
 The  black-shaded histograms in the figures (a) and (b) are
 calculated mass yields using the microscopic and the macroscopic densities,
 respectively. The solid-line histograms are  experimental data taken from
 Ref.~\cite{FmData}.} 
\end{figure}
 
Figure 6 shows the macroscopic and microscopic mass-yield curves.
The same data are shown as histograms in Fig.~7, so that we can compare
directly to the experimental data, which are given in histogram
form~\cite{FmData}. In transforming to the  histograms, we
 calculate the average of integrated values of the mass-yield curve for the
 4 u interval used in the presentation of the experimental data~\cite{FmData}.  
 The solid and the dashed lines in Fig.~6 correspond to the microscopic and the
 macroscopic densities, respectively.
 The calculated peak value is consistent with the experimental one,
 although a one-dimensional model could have a tendency to overestimate the peak
 value of the mass-yield curve~\cite{LIC73}. On the other hand, it is
 clearly seen that the result of the macroscopic density is not
 in agreement with the extremely narrow peak of 
 the experimental mass-yield curve.  That is, the 
  single-particle configuration, and specifically
the large shell gap, is the source of  the
 extremely narrow FWHM of the fragment mass-yield curve. 
 We thus expect that the FWHM of the compact symmetric component is 3.6 u.
 
 In order to obtain the whole mass distribution, it would be necessary
 to clarify valley structures for all fission paths and superpose their
 contributions. However, we could not employ our scission-point
model to the other valleys, because it was not possible to obtain unique scission points
 for the asymmetric and elongated symmetric paths. For those paths  the separating
 ridge vanishes before the scission configurations, indicating that these
 components must be modeled in a more complex dynamical approach.
 
One may ask how the features we studied here manifest themselves
in self-consistent mean-field calculations which have also studied $^{258}$Fm.
In those models the potential and microscopic densities are ``self-consistent''.
One could anticipate that perhaps no self-consistent solutions exist for
octupole constraints corresponding to the range 0 to 0.18 of $\alpha_{\rm g}$
in Fig.~2. And when a solution exists there would be a large increase in
energy. If these expectations do occur in self-consistent models
this behavior would tend to very much restrict the fluctuations
in the mass-yield curve, just like in our studies here.
However, in the papers \cite{Bonneau06,Dubray08} we find no results that 
shed light on precisely these issues. But the calculated barrier
versus a quadrupole constraint in \cite{Bonneau06} is very similar to the barrier
obtained here. 
Recently \cite{Dubray08} looked at correlations between various fission
fragment properties of
Fm isotopes. These calculations do not clearly identify any fragment
shell effect on mass yield widths. This is probably because
they were not specifically designed to study such a possibility.
 In Ref.~\cite{Dubray08},  the Hartree-Fock-Bogoliubov
 method is used to calculate  potential-energy surfaces of Fm isotopes along
 a scission line.  The results for the  $^{258}$Fm potential energy shown in  Fig.~7 have
 quite different shape compared to our Fig.~4(c). In particular the potential
 shows a very shallow minimum centered at  symmetric division. 
The comparisons of calculated total kinetic energies to data
is not very convincing.
The substantially different results we obtain are likely due
to the very different designs of the two studies, not to
the differences between self-consistent and non-selfconsistent models. 
In our case we study mass oscillations
near the exit point of compact scission. In the HFB study energy partitioning
along the entire scission line is the mechanism governing the fragment properties.
Since that mechanism shows substantial differences with respect to measured
data, we feel it is more appropriate to 
describe origin of the narrow mass distribution
 in terms of the large shell gap at the barrier exit point, as we do here.

 In summary, we have presented potential-energy curves of $^{258}$Fm at
 scission  as  functions of both macroscopic and  microscopic fragment mass densities.
 We calculated the zero-point vibration corresponding to this potential-energy
 curve by solving a  one-dimensional Schr\"odinger equation with a
 coordinate-dependent inertial mass based on the cranking model.
 An important point in the calculation is that the fragment masses are defined by
 the  single-particle 
 wave functions, rather than the macroscopic potential volumes.
 We have shown that the mass numbers of two nascent fragments strongly
 depend on the single-particle configurations. In particular, the
 proton number of the fission fragments originating from the compact
 symmetric valley for 
 $^{258}$Fm is strongly constrained  to $Z=50$ due to the large shell gap.
 The calculated mass-yield curve is consistent 
 with the extremely narrow experimental mass yield curve.
 We obtain that the FWHM of the fission fragments originating from the
 compact symmetric valley is 3.6 u.
 For $^{258}$Fm, it would be interesting to measure the ratio of protons
 to neutrons on the mass yield curve, because
the neutron distribution may be wider than the proton distribution, since the mean fragment neutron
number is not magic. This would be a very strong test of the mechanism behind
the narrow mass distribution. 
  
\begin{acknowledgments}
 TI is grateful for the Special Postdoctoral
 Researcher Program in RIKEN.
 The numerical calculations have been performed at the
 RSCC system, RIKEN.
PM would like to acknowledge that this work was carried out under the auspices of the National Nuclear
Security Administration of the U.S. Department of Energy at Los Alamos
National Laboratory under Contract No. DE-AC52-06NA25396 
and  was also supported by  a travel grant to 
JUSTIPEN (Japan-U.S. Theory Institute for Physics with Exotic Nuclei)
under grant number DE-FG02-06ER41407 (U. Tennessee).  

\end{acknowledgments}

\end{document}